# Harmonics Generation in Ultra-Thin Nanofilms Irradiated by Intense Nonrelativistic Short Laser Pulse.

Ph. Korneev,
NRNU MEPhI
korneev@theor.mephi.ru

## Abstract

The problem of harmonics generation in nanotargets is considered at the range of parameters (a nanotarget diameter and a pump laser intensity) when the oscillation amplitude of an electron in a target is much larger than the target width. Electron motion in charged nanotargets in the presence of a laser field of different (non-relativistic) strength is considered. It is demonstarted that for lasers of infrared frequencies clusters do not posess strong enough potential to bound electrons with large oscillation amplitudes. Opposite to clusters, nanofilms were found to be very perspective targets in the problem considered. A simple analytic model and molecular dynamic simulations showed increased harmonics generation when the oscillation amplitude of electrons in a film becomes much larger, than the film width. Different regimes of generation are briefly discussed.

## 1 Introduction

Since the great break-through in laser technologies made a few tens years ago [1], high-intensity interaction of laser fields with matter is still being of a great interest. Because of the high laser intensity a plasma is formed on the front of such pulses during the irradiation of any target. Such a plasma may posess different properties depending on laser intensity and carrying frequency, irradiated target, etc. A lot of new and promising phenomena in plasmas were found in the last decade: high harmonic generation on a plasma surface [2, 3], electrons acceleration by a laser wake-field [4] and in the bubble regime [5, 6], protons acceleration [7–9]. Also a lot of interesting effects were observed and theoretically described in the case when an intense laser interacts with nanotargets (nanoclustrers, wires, and thin films). One of those is harmonics generation, which was observed experimentally both in films and nanoclusters (see, for expamle, [10–13]). As far as we know, in the case of films, usually experiments were held with film thickness more than 100 nm, and relativistic intensities were used. That was because if the intensity is not enough high, after an ionization the overdense plasma permits further penetration of the laser radiation into the film. Relativistic effects such as 'light pressure' allow to go through this. However, nanotargets of the 1-10 nm scale (noble gas clusters) were found to be very efficient absorbents of light even at atomic intensities [14]. That is the reason why the nonlinear responce such as harmonics generation of such a medium was so interesting to study.

High harmonics, observed in experiments on a laser-nanoclaster interaction at atomic and subatomic intensities [11–13] should probably be related to the 'atomic' mechanism, in which electrons recombine to their parent or neighbor ions [15–17]. At higher, but nonrelativistic intensities only several lowest harmonics were observed [18, 19]. The origin of the lowest harmonics observed in experiments [18, 19] can be understood in terms of the nanoplasma model. According to that, an interaction begins with a rapid inner and outer ionization on the front of the laser pulse with creating of a nanoplasma, which is then heated [20–24], interacts with the laser field on a femtosecond scale, and finally the Coulomb explosion occurs because of the total positive charge of the target [25]. When a laser intensity is high enough, the nanoplasma is hot and collisionless, which allows to describe it with the Vlasov equations as a gas in a self-consistent field. Within this approach harmonics generation process was theoretically described in [26–28]. It was also found that the lowest harmonics of the laser frequency may be amplified in a resonance with the Mie frequency [29], which changes during cluster expansion. Moreover, higher harmonics were also found to be generated, but effectively they are suppressed because of stochastic electron motion in a heated cluster [30–32]. This is the reason why only the lowest harmonics were observed in the experiments [18, 19].

The reason of stochastic behavior of electrons showed in [30, 31] is well known as a nonintegrability (see, for example, [33]) in the case when an external laser field and a quasistatic self-consistent potential in a nanobody are both of the same order of magnitude.

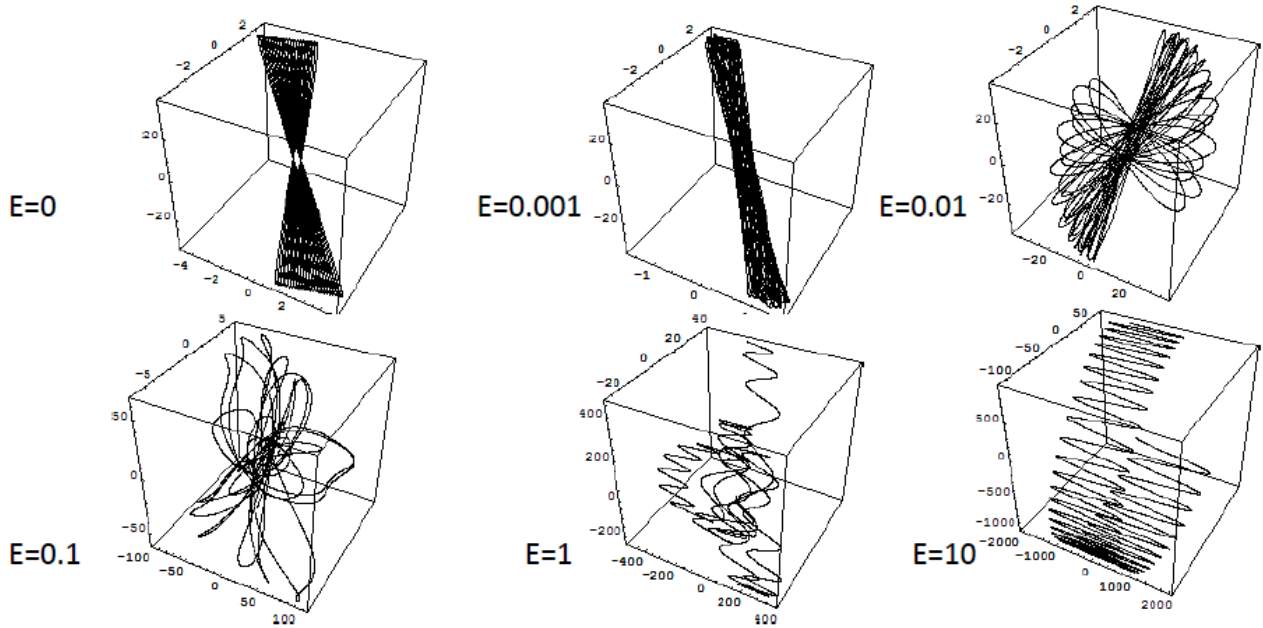

Figure 1: Different trajectories for electron in the charged cluster under the action of laser field (4) of different intensities. When an incident field is much less or much greater than the potential of a charged cluster, trajectories are regular (1,2,3,6 panels). At the intermediate region stochastic dynamics develops (4,5 panels). See text for further explanations.

The relation between the self-consistent field and the incident laser field may be defined as a parameter

$$\xi = \frac{a}{R} = \frac{e\mathcal{E}}{m_e \omega^2 R}, \qquad (1)$$

where $a = e\mathcal{E}/m_e\omega^2$ is the electron (with mass $m_e$ and charge e) oscillation amplitude in the external harmonic laser field $\mathcal{E}$ with frequency $\omega$, and $R$ is the radius (width) of the considered nanotarget (cluster, film, etc.) If $\xi \ll 1$, then the nonlinearity in the system is small, and only the lowest harmonics of the incident field are generated [26–29]. In this case the standard perturbation theory works fine. When $\xi$ grows with the laser intensity the nonlinearity of a nanosystem encreases and becomes considerable at $\xi \sim 1$, but the nonintegrability mentioned above does not allow the coherent addition of harmonics generated by different electrons. That is a question if an increase of the laser intensity up to values when $\xi \gg 1$, cause an approximate integrability to come back and allows to treat the problem in terms of perturbation theory with a small parameter $\xi^{-1} \ll 1$. In this case the self-consistent potential of the charged nanobody is a perturbation while an incident laser field plays the main role, nonlinearity however is great in this case so that as a result an extended harmonic spectrum may be generated by the nanocluster.

A vivid demonstration (see figure 1) of the different laser-cluster interaction regimes defined by the parameter (1) may be given by a direct integration of Newtonian equations for an electron in the potential of a charged sphere, which corresponds to a small one time outer ionized cluster with radius 1 nm. For different values of $\xi$ trajectories are shown on the figure 1. When the laser field is off ($\mathcal{E} = 0$, the first panel on the figure 1), an electron oscillates according to its initial condition: this is fully integrable problem since the potential is static and center-symmetric. Increasing of the laser field amplitude ($\mathcal{E} = 0.001$ for the second panel and $\mathcal{E} = 0.01$ for the third panel on the figure 1, all values are given in atomic units) leads to more complex trajectories. This situation corresponds to $\xi \ll 1$. The system is not fully integrable in this case, however, perturbation theory may be helpful: it is seen from the figure 1 that trajectories are quite regular. Further increasing of the laser field strength gives a birth to unstable trajectories ($\mathcal{E} = 0.1$, fourth panel, and $\mathcal{E} = 1$, fifth panel on the figure 1), the system is not integrable at all, and demonstrates a stochastic behavior. In this case $\xi \sim 1$. When the laser field strength is more increased, the system enters the region where $\xi \gg 1$, and, as it is seen from the figure 1, sixth panel, for $\mathcal{E} = 10$, the electron motion becomes regular again. The system is not fully integrable, the same as in the case $\xi \ll 1$, but again as in that case it may be desribed with the use of perturbation theory.

We examined this situation ($\xi \gg 1$) for clusters and found that in the realistic range of parameters it is very difficult not to say impossible to find a system which can produce harmonics. To come to the region $\xi \gg 1$ the oscillation amplitude of an electron should be large compared to a cluster size. From the simple analysis it is seen that an incident laser field should be of the order of several atomic units, which is also seen from figure 1, sixth panel.

For a Ti:Sa laser this condition may be fulfilled only for small clusters (less or of the order of few hundreds of atoms), but the self-consistent potential in this case is too weak to confine electrons near the cluster even for few periods of the laser field (the same behavior was reported in [20]). But without action of the cluster potential the electron would not radiate. A small time of life of such small clusters even agravates the situation. Some of the described problems may be solved by changing the geometry of the target, and considering nanofims instead of nanoclusters. The self-consistent potential of the film is large on the distance of the order of a charged area diameter, which may prevent the continuos outer ionization even for thin films.

In case of nanofilms exhaustive investigation of the nonlinear responce of thin films to p-polarized laser pulses in the case $\xi \lesssim 1$ was presented recently [34]. Hydrodynamic approximation was used to describe a nanofilm behavior in laser fields of moderate intensities ($10^{14}$ .. $10^{16} \text{W/cm}^2$). It was found that the film boundaries play the decisive role in the nonlinear response of the whole system. The thickness of films was considered much larger than 1nm. In the present work we consider the different case of utrathin nanofilms with the thickness $\sim 1$ nm, and a bit higher intensities than in [34][1], so that

$$\xi \gg 1 \tag{2}$$

and a p-polarized laser field. To fulfil the condition (2) the width of considered nanofilms should be as small as few nanometers and intensity $I \gtrsim 10^{17} \text{W/cm}^2$. We stand in a nonrelativistic domain for laser fileds, to avoid relativistic effects such as a light pressure.

The Letter is organized as follows: in the next section we define parameters of interaction, a model and a qualitative analytical description. After that we present molecular-dynamic results for short and longer pulses, and finally conclude.

## 2  Statement of the problem

To obtain an effective harmonics generation several conditions should be fulfilled simultaneously. First, different dipoles (electrons) should oscillate approximately in phase. The difference between the phases of the order of $2\pi/\Omega$, where $\Omega = k\omega$ is the frequency of a generated harmonic destroys coherence and may define harmonics cutoff. Second, oscillations should be nonlinear, which means in our case that the amplitude of oscillations should not be small compared to the system width. For the condition of a phase-matching it would be better if the system is only slowly changing in time during the interaction process. For this we consider films of heavy materials, such as metals (Au, Pt, etc.) These metals are also known to be good for nanofilms production. First of all we analyzed an electron dynamics in the potential of a charged film, in a way we do it earlier for a charged cluster. To avoid unphysical singularities we modeled the charged film with a smoothed force, which gives correct limiting cases for a charged film with the thickness $R$

$$F = 2\pi e \sigma \frac{x}{\sqrt{R^2 + x^2}} = 2\pi e Z n_i R \frac{x}{\sqrt{R^2 + x^2}}, \qquad (3)$$

where e is the electron charge, $R$ is the film width, $Z$ is the average ion charge, and $n_i$ is the atomic density of the matter of the film, $x$ is the coordinate of an electron. The laser field is defined as

$$\mathcal{E}(t) = \mathcal{E} \cos \omega t \exp\left[-\left(\frac{t-t_0}{0.5 t_0}\right)^2\right], \qquad (4)$$

where $\omega$ is the laser frequency which is throuout the letter is 0.057a.u. = 1.5 eV, $\mathcal{E}$ is the electric field amplitude, $t_0$ defines a duration of the pulse. For the analysis below we used $\mathcal{E} = 0.5..2.5$a.u., $R = 20$a.u. (1 nm), $t_0 = 2000$a.u.($\sim 100$fsec), outer ionization degree per one ion $Z = 1$, which value gives for the surface charge of the film $\sigma = 0.175$a.u., taking into account atomic density for Au $n_i = 0.00874$a.u.$^{-3}$. We considered then different values of $\mathcal{E}$. Solutions (trajectories) of the Newtonian equations of motion, with zero initial conditions unless otherwise stated are presented on figure 2. When the filed amplitude $\mathcal{E}$ is small (0.4a.u.) comparing to the electrostatic force (3), which is $\sim 0.7$a.u. in our case, then oscillations take place at the bottom of the potential of the film (first panel on the figure 2). Increasing of the laser intensity leads to absorption of laser energy. It begins when the oscillation amplitude becomes equal to the film width $R$, i.e. when $\xi \approx 1$, which corresponds to $\mathcal{E} = 0.51$a.u.. As a result, when the pulse is gone, the electrons oscillate in the self-consistent potential (second panel on the figure 2). To generate radiation different electrons should oscillate in phase. This only may be realized for some part of electrons. If this is possible, then oscillations on the order of the plasma frequency of the system would survive. Further increasing of the laser field amplitude changes the electron dynamics dramatically. For $\mathcal{E} = 0.6$a.u. the electron goes through a nonlinear resonance, effectively absorbs energy, and after the laser pulse is gone, oscillates with an amplitude larger than it would be in the laser field without film potential (3) (note that on the 3-d..6-th panels on the figure 2 the scale is 100 times larger than on two first ones). This is a well known behavior of a nonlinear system, when it goes to the nonintegrable domain. In this domain trajectories are very sensitive to initial conditions. Such a behavior is not changed untill $\mathcal{E} \sim 2$a.u. (see panel 4 on the figure 2), when finally the system enters the region of parameters, where trajectories look similar for wide range of initial conditions. This corresponds to $\xi \gg 1$, and $\mathcal{E} \gtrsim 2$a.u. On panels 5-th and 6-th on figure 2 for $\mathcal{E} = 2.5$a.u. two trajectories for initial conditions on the opposite edges of a film ($x(t = 0) = -10$a.u. and $x(t_0 = 0) = 10$a.u.) are shown, and this two trajectories hardly differ one from the other.

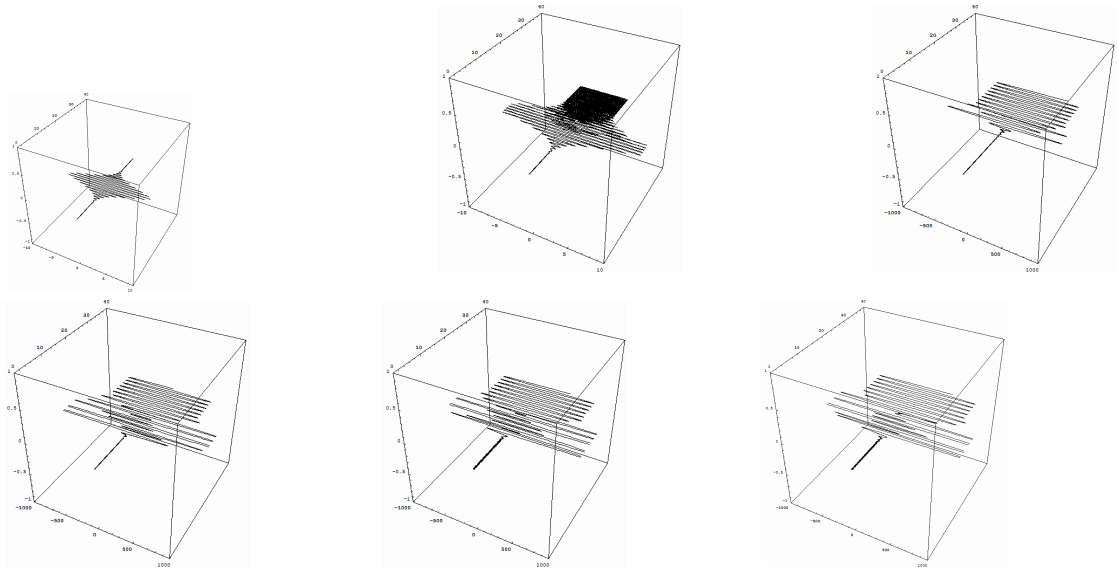

Figure 2: Different trajectories of an electron in the charged film under the action of laser field (4) of different intensities, unwraped on time. When the laser field intensity is small, the electron oscillates at the bottom of the potential (panel 1). At higher intensity it comes through the nonlinear resonance, and oscillates after the pulse is gone (2,3, panels). At the $\xi \gg 1$ regime electrons trajectories for different initial conditions are almost the same (panel 5,6), which indicates that the motion of the electron is regular. See the text for further explanations.

---

[1] Free-standing films of up to such extremely small thickness may be available at the modern level of technology [35]. Note, that such films of the 5 nm width were already produced more than ten years ago [36, 37].

When the condition (2) fulfils perturbation theory with respect to parameter $1/\xi$ may be applied for an analitical analysis of the system. This however is not a well defined perturbation theory: $1/\xi$ is not really a parameter in it. The reason is that in a potential of a charged plane (as well as in any potential which is proportional to a some power of a distance) there is no characteristic length. The film width does not play a role on a large distance. We apply a perturbation theory here after the analysis of trajectories presented earlier, which shows that electron's motion is regular for a large field, and examine the results with simulations in the next section.

For the perturbation theory series we now consider only one electron in the self-consistent field of the film of zero width, and the static potential, which is a good approximation on a time scale less than the time of the Coulomb explosion [26, 38]:

$$\tau_{\mathrm{ion}} \sim \sqrt{\frac{M_{\mathrm{i}}}{e^2 n_{\mathrm{i}} Z^2 \kappa}}, \qquad (5)$$

where $M_i$ is the ion mass, $\kappa$ is the ionization degree. Let the unperturbed trajectory of an electron is $x_0(t)$, which is defined by the Newtonian equation with the laser field only:

$$m_e \ddot{x}_0 = e \mathcal{E} \cos \omega t, \qquad (6)$$

with some initial conditions and a perturbation to the right-hand of (6) is

$$F(x) = 2\pi \sigma \mathrm{e} \times \mathrm{sign}[x], \qquad (7)$$

where $\mathrm{sign}[x] = x/|x|$ is the signum of $x$. In the first approximation we then have

$$m_e \ddot{x}_1 = -2\pi \sigma \mathrm{e} \times \mathrm{sign}\left[-\frac{\mathrm{e}E}{m_e \omega^2} \cos \omega t\right]. \qquad (8)$$

For dipole radiation the spectral distribution of the irradiated energy is given by [39]

$$d\mathcal{W}_\Omega = \frac{4 \mathrm{e}^2}{3 c^3} |\ddot{x}_\Omega|^2 \frac{d\Omega}{2\pi} \qquad (9)$$

where we should put from (8)

$$\ddot{x}_\Omega = \frac{2\pi\sigma e}{m_e} \int_{-\infty}^{\infty} \exp[i\Omega t]\text{sign}\left[-\frac{eE}{m_e\omega^2}\cos\omega t\right] dt. \quad (10)$$

Evaluation of (10) is simple due to the presence of the function sign$[x]$. We note, that its derivative is the $\delta$-function, and integrate (10) by parts. Then

$$|\ddot{x}_\Omega| = \sum_{n=-\infty}^{\infty} \frac{64\pi^2\sigma^2 e^2 T_\infty}{m_e^2(2n+1)^2} \delta\left(\Omega - (2n+1)\omega\right), \quad (11)$$

where we used a well known method of substitution

$$\delta^2(\omega - \omega_n) = \delta(0) \cdot \delta(\omega - \omega_n), \quad \delta(0) = \lim_{T\to\infty} \frac{T}{2\pi},$$

and then $T_\infty$ is a full time of the observation. This gives for the $2n+1-$th harmonic intensity

$$I_{\Omega=(2n+1)\omega} \equiv I_n = \frac{256\pi^2\sigma^2 e^4}{3c^3 m_e^2} \frac{1}{(2n+1)^2}. \quad (12)$$

The spectrum slope (12) can not extend up to infinity. This should be, if we would have ideal case (7) with singularity at zero point. In reality this singularity changes to a smooth (close to parabolic) potential on a distance of the film width. One then can estimate the cutoff frequency as a ratio between the oscillation amplitude[2] and the film width:

$$\omega_{cut} \sim \frac{e(\mathcal{E} - 2\pi e\sigma)/m_e\omega^2}{R}. \quad (13)$$

For the considered parameters ($\mathcal{E} = 2$ a.u., gold, $Z = 1$) this gives $\omega_{cut} \sim 15$ for the 1 nm-width film. When ionic sceleton expands $R$ increases and $\omega_{cut}$ drops.

The result (12) is obtained in a simplified fashion, that means that it is questionable at some points. First of all (i), the perturbation theory is based on an assumption of regularity of electrons motion when $\xi \gg 1$, as discussed above. Second (ii), we considered only one electron, thus did not take into account phase relations between different electrons, which are very important for collective radiation. And (iii) we considered the self-consistent potential to be a potential of a charged film, i.e. did not take into account collective effects. The plasma oscillation frequency for the parameters considerd has in the unperturbed film a value of about 1.5 a.u., which is about 30 times greater, than the carrying laser frequency $\omega$. When electrons are being heated, its density and the plasma frequency decrease, as it is seen from the simulations below, for different values from 3 to 15 times the laser frequency. This fact makes the estimation (13) reasonable only up to that final value of the plasma frequency. To check the simple theory (12) we presented numerical simulations of the irradiated ultra-thin nanofilm. The system considered in the present Letter is a non-relativistic one, and because of the ultra-low thickness of the cosidered films it contains not really a macroscopic number of particles (ions and electrons) per simulation volume. This makes possible to make a numerical experiment with MD simulations. Propagation effects such as the screening effect inside the film for such thin films are correctly described by MD, as far as thickness is much less than the anomalous skin-depth, the plasma wave length, etc. The skin depth is defined by

$$L_{\text{skin}}^{\text{a}} = \frac{c}{\omega_p}, \quad (14)$$

with the plasma frequency

$$\omega_p = \sqrt{\frac{4\pi n_e e^2}{m_e}}, \quad (15)$$

where $n_e$ is the electron density, and for the inner ionization degree $Z$ it has the order of $n_e \sim Zn_i$ before the sighificant outer ionization proceeds. Even if $Z \approx 8$, the plasma frequency is $\omega_p \sim 1 a.u.$, and the skin-depth is $L_{\text{skin}} \sim 130 a.u. = 6.5 nm$, which is much greater than the considered film width.

We used LAMMPS[3] [40] for MD simulations presented. We started in the simulations with the prepared ionized films: ions were placed as in FCC lattice in gold, electrons were distributed randomly in the volume of the film with zero velocity. This would be replaced by ionization under the action of the field in future work, at the moment we just test the possibility of harmonics generation in nanofilms, and for that goal our approach is sufficient. Charged particles (electrons and ions) were interacting by the Coulomb forces, calculated with PPPM [41] option, which allows to get practically infinite range of interaction and to model an infinite film.

---

[2] We took here into account the shielding by the charged film
[3] See http://lammps.sandia.gov.

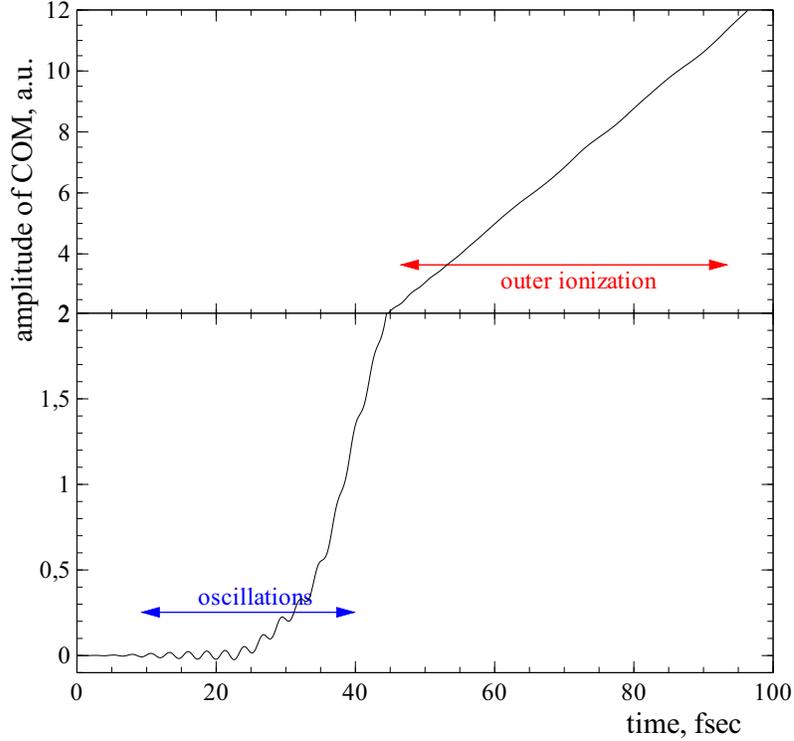

Figure 3: Typical curve of the total diplole moment, for 1 time per an ion outerly ionized film with the filmwidth 1 nm, the pump field strength 2 a.u., and the pulse duration 25 fsec. The total time of MD integration is 100 fsec. Two regions are clearly seen on the figure: the oscillations and the evaporation of electrons (outer ionization).

## 3 Results, analysis, discussion

We performed plenty of runs for different parameters and found that there are common features for very different situations. We found that a part of the electrons is more strongly bound to the film and oscillated within a small region near the bottom of the self-consistent potential, while another part comes through the nonlinear resonance (see figure 2) and oscillates with large amplitudes compared to the film width. The later part is namely that for which the perturbation theory (6-12) presented above, and these electrons evidently produce the main part of the harmonic spectrum. The common features in the motion of the electrons with large oscillation amplitudes may be visually demonstarated with the time-dependence of the total dipole moment of the intense laser irradiated film. The typical total dipole moment behaviour is shown on figure 3. When the laser is turned on, first the dipole moment oscillates, which means that the electrons oscillate within some distance around the film, but closer to the end of the pulse (for the given parameters $\mathcal{E} = 2$ a.u., $R = 1$ nm it turns out to be after about 15 fsec) the electrons may leave the potential well of the film and go to infinity. This is not possible in an ideal infinitely large thin film, but in reality the charged area is of the order of a laser focusing spot $(2 - 10\mu\text{m})$. We considered electrons to be free when they come out from the film on the distance of $10\mu\text{m}$. These electrons form an increasing tail in the dipole moment, which goes to infinity in the infinite future. The time moment of switching from oscillations to evaporation $t_{sw}$ defines two regimes of harmonics generation. Before this moment the electron cloud is more confined, and the electrons oscillate more in phase. After the moment $t_{sw}$ the electronic cloud is destroyed, and some of electrons go to infinity. So it is possible to distinguish between short pulses when the electrons are still in the potential of the film and the electron cloud is confined, and long ones, when some part of electrons leaves the film potential and the cloud is destroyed. We thus divided further consideration for two parts: 'short pulses' and 'longer pulses'. In further MD simulations we considered films of gold, and we use parameters of metallic gold, which has FCC lattice, with the length $407.82 pm$, the volume of one cell then is $457.757 \text{a.u.}^3$, in one elementary cell there are 4 atoms, and then atomic density is $0.00874 \text{a.u.}^{-3}$. The given numbers are important because they define the electronic plasma frequency, which is for this atomic density and assuming ionization of 8 electrons, is $\omega_p \sim 30\omega$. Note again, that this is the initial plasma frequency, which decreases while more and more electrons escape from the film potential.

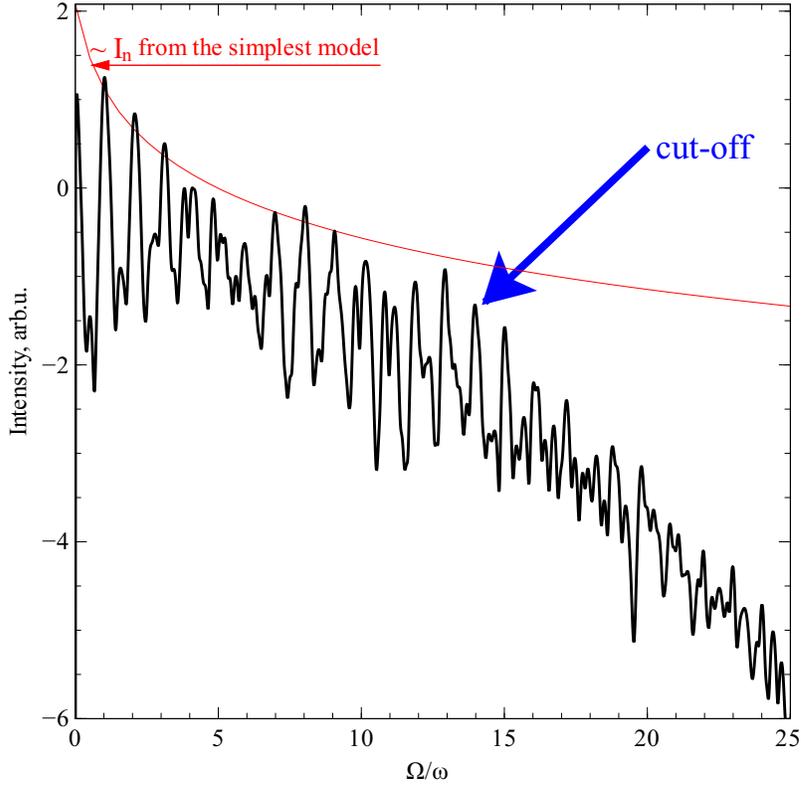

Figure 4: Harmonic spectrum for the 1 nm width 1 time per ion outerly ionized, the pump field is 2 a.u., the pulse duration is 15fsec, MD integration time is 60 fsec. Red curve is the roll-off from the simple theory (12) with a multiplication factor as an adjustive parameter. The blue arrow shows cutoff from (13). Both roll-off law and cutoff frequency are in a good agreement with the result of the simplest model (12). The difference is in the presence of even harmonics. See the text for the explanation.

### 3.1 Short pulses

In this subsection we consider only short pulses with the time duration less than $t_{sw}$, which allows to expect that the simple theory (6-12) should work better because the electron cloud is more confined. Considering the dynamics of electrons in the potential of a charged film, we can find from [42], that the absorption rate (energy, absorbed per unit time) may be estimated for an electron as $\sim \epsilon^{-3/2} \times \log \epsilon$. A decrease of energy absorbed by the collisionless nanoplasma per time unit comes from the decrease of the number of collisions with the potential boundaries when electron's energy increases [42]. This rate gives the average energy of the electrons in a film versus time. When the pulse duration is small, the electrons can not absorb enough energy to come to high levels and to decrease strongly the electronic density. So collective effects would not influence critically the lower part of the emission spectrum where $\Omega \ll \omega_p$. Moreover, when the pulse is small and the electrons are cold in that sense that their relative (thermal) velocities are small. In this case they all oscillate more as a one confined cloud, which means that we are on the left part 'oscillations' on figure 3.

The simulation was ran for totally 60 fsec with the film of the width 1nm, the laser field of 2 a.u., and the laser pulse duration of 15 fsec. We calculated the dipole moment of the system and its second derivative to obtain the radiation spectrum. The square of the second derivative of the dipole moment is shown on figure 4. Harmonics are nicely seen on the figure. The slope curve is rather flat. To compare the result (12) for the simple model presented earlier with the data from simulations we plot also on figure 4 a red curve, which indicates the prediction of the simple theory with one adjustable parameter – the total intensity. It is seen, that the agreement of the roll-off law is qualitatively very good. The most important difference comes out the fact that on the data curve we see the presence of even harmoncis, while there should not be any in the simple theory. Indeed even harmonics should not appear in the case of a symmetric potential, which we have in the simple model (6 - 12). We analized the behaviour of different electrons and have found that there is a part of them which is more bounded in the bottom of the potential field of the film, and the motion of these electrons has not always the same phase as the part of electrons with large oscillation amplitudes has. This in different words may be told as a compressibility of an electron cloud. The part of the electrons at the bottom of the film potential effectively destroys the symmetry and even harmonics then may appear in the spectrum. However, the potential at large distances compared to the film width does not change dramatically, and total roll-off of the spectrum is reproduced by the simple model (6 - 12).

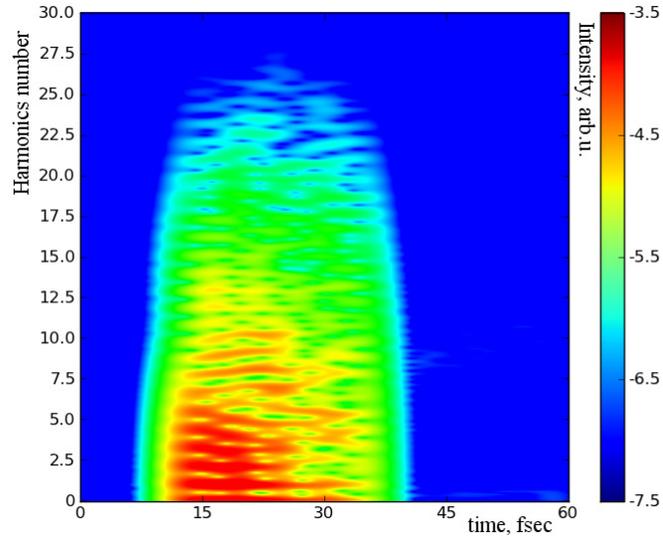

Figure 5: Time-resolved harmonic spectrum for the 1 nm width 1 time per ion outerly ionized, the pump field is 2 a.u., the pulse duration is 15fsec, MD integration time is 60 fsec. No collective effects are seen as there is no a significant change of the spectrum during the action of the pump wave.

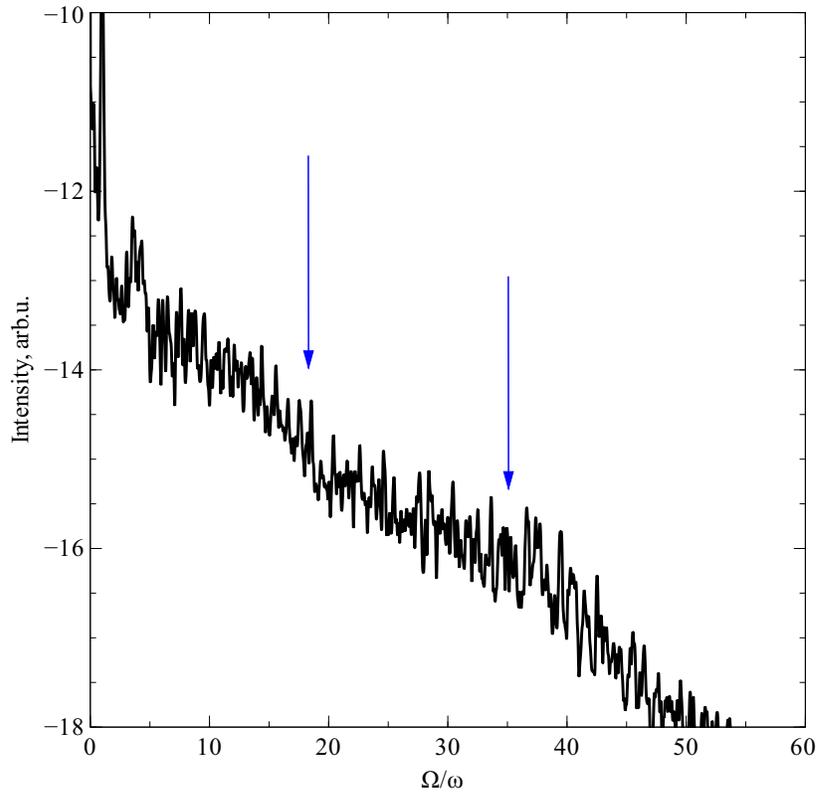

Figure 6: Harmonic spectra for the irradiated film of 1 nm width, the field strength is 2 a.u., the pulse duration is 25 fsec, total MD integration time is 100 fsec. The fifths harmonics enhancement, the first (decreased plasma frequency) and the second cut-offs are seen.

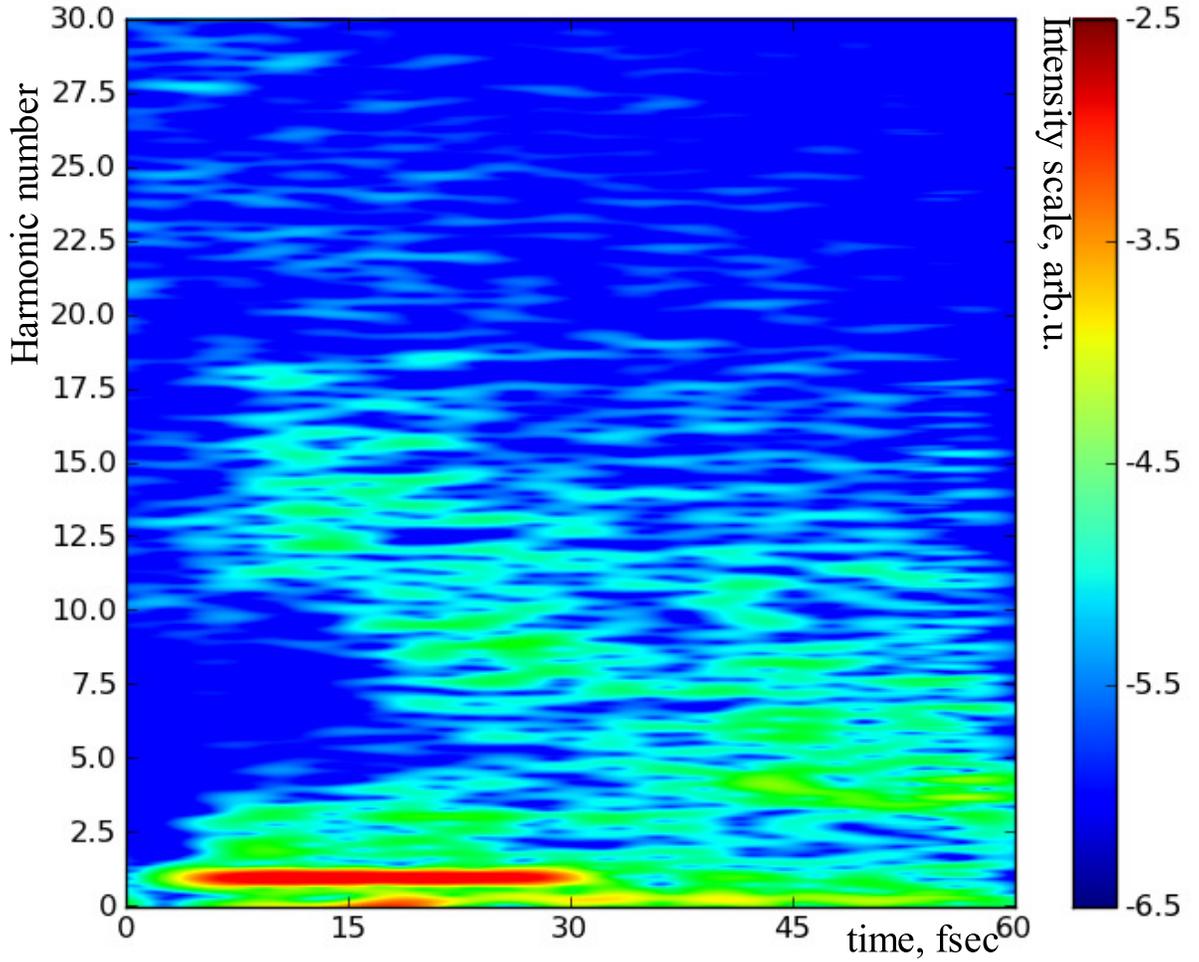

Figure 7: The same as on figure 6, but time-resolved. The plasma frequency decrease and residual plasma oscilations on the fifth harmonic are seen.

On the data we see also some enhancement after the fifth harmonic. We estimated the plasma frequency for the number of electrons near the film and found that it is much higher, so that the enhancement can not be attributed to plasma oscillations unless it decreases down to the order of 5-th harmonics of the laser. To clear this out we plotted the time-resolved spectrum for this simulation, which is shown on figure 5. It is seen from the figure, that the enhancement is not developed in time, so it can not be attributed to the change of plasma frequency. Moreover, we can state here that no collective effects are developed when the laser pulse is as short as 15 fsec. This means that our assuption concerning applicability of the simple model at short pulses is correct.

## 3.2 Longer pulses

For the case of longer pulses, as we expect, the decreasing of plasma frequency should play a role in the process of harmonics generation. We consider now the same films of 1 nm width, but the pulse duration (FWHM) is encreased up to 30 fsec. We consider both neutral and charged films at different intensities. The interaction process was studied

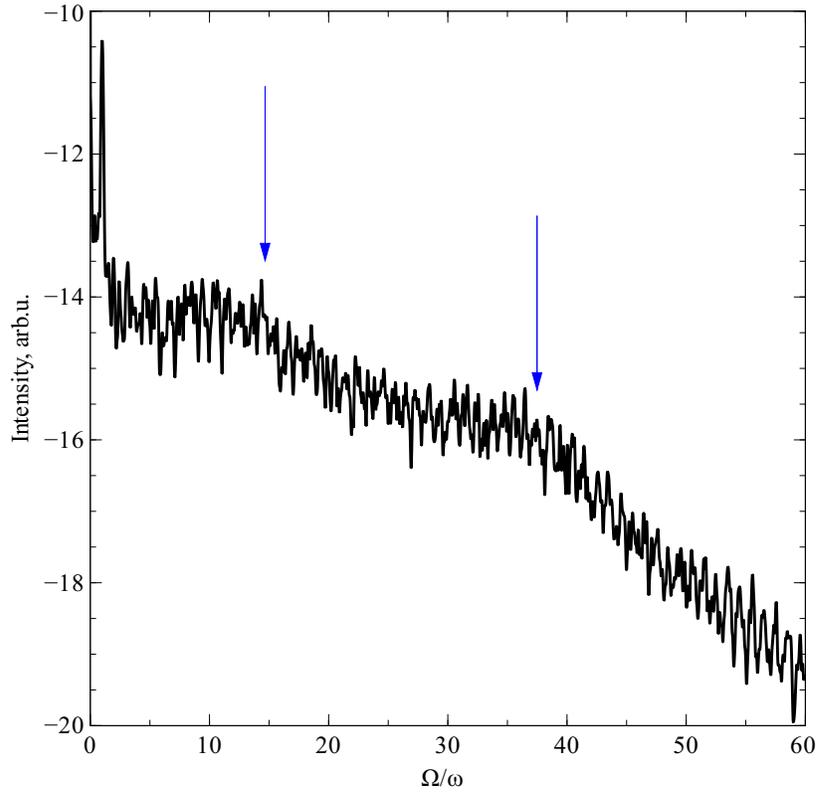

Figure 8: Harmonic spectra for the irradiated film of 1 nm width, the field strength is 1 a.u., the pulse duration is 25 fsec, total MD integration time is 100 fsec. Fifth harmonics enhancement, the first (decreased plasma frequency) and the second cut-offs are seen.

during 100 fsec to look for the residual oscillations, if any.

On figures 6 and 7 the harmonic spectrum and the time-resolved harmonic spectrum are shown for the case of the laser field with $\mathcal{E} = 2 a.u.$. Comparing this two figures we can see that the fifth harmonic is enhanced when the plasma frequency decreases to that value (at the end of the laser pulse), and forms residual osciations which are not effected by the laser pulse because it ends and should stop by the Coulomb explosion of the film. We see also that the plasma oscillations are developed on the time scale greater than 15 fsec, which one more time proves that different regimes correspond to different laser pulse durations. There are two different cut-offs seen of figure 6. The lower one corresponds to the initial value of the plasma frequency, and this may be concluded from the time-resolved spectrum on figure 7. The higher cut-off is a puzzle, though it is practically too low to be important, but there is no such a frequency in the system. The only possibility we see is the second harmonics of the plasma frequency. We would return to the question of the plasma-frequency harmonics in the last section.

On figures 8 and 9 the harmonic spectrum and the time-resolved harmonic spectrum are shown for the case of $\mathcal{E} = 1 a.u.$. All the features of the spectra are common to the previous figures 6 and 7. The difference is in the total plasma frequency excitations amplitude and the absence of the resonance at lower frequencies are understood just from the lower laser strength. From one hand the electron cloud is more dense, because electrons are not so much heated, from the other a small strength of the pump wave is not able to excite the higher frequency oscillations than in the prevoius case for $\mathcal{E} = 2 a.u.$.

The next simulation we ran was for a two times outerly ionized film. This situation is probably more realistic since after ionization electrons have an average energy of the order of the ponderomotive energy in the field, which may be enough to escape from the potential of the film. The outer ionization forms a large positive charge of the film. It strongly bounds the electrons and prevents a rapid decrease of the electronic density. As a result, the plasma frequency decrease is much slower and even stops at higher values compared to the previously considered case of neutral films. At that values of the plasma frequencies the laser pulse resonantly enhances 7-th and 9-th harmonics, which are closest to the plasma frequency at the moment. This ampification is very strong and the 7-th harmonics intensity on figure 10 is seen to be only one and a half orders of magnitude less than the laser frequency $\omega$. The described effect is a resonant enhancement of several harmonics in the film looks very common to the same effect in clusters, but has quite different nature. In clusters this happens mostly when the Mie frequency decreases because of the Coulomb explosion of a cluster [43], [44], so it is difficult to control the process. The ionic core of a charged cluster is expanding quickly already when the resonance occurs and so the possible time of resonant oscillations is then much smaller than the Coulomb explosion time. In films the decreasing of the plasma frequency is a more controllable process which is defined by the rate of electron heating (laser intensity, frequency, polarization, etc.) and the total charge of the film (a degree of the outer ionization). A resonant condition may be achieved in a few femtoseconds after the laser pulse is turned on and be fulfilled until the Coulomb destruction of the film. This means that the time for resonant oscillations may be as large as the time of the Coulomb explosion (hundreds of femtoseconds and more).

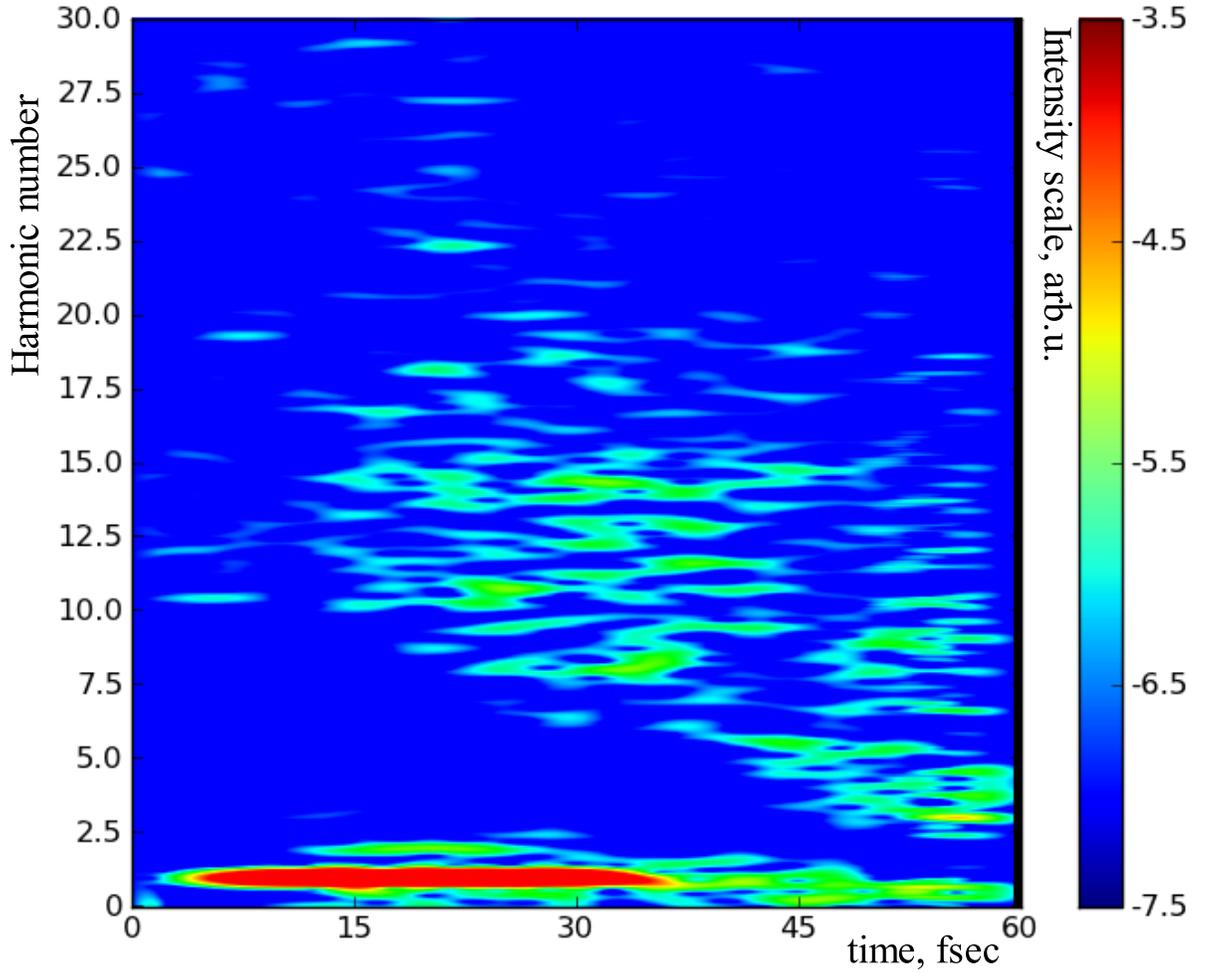

Figure 9: The same as on figure 8, but time-resolved.

We also considered 4 times outerly ionized film, but did not find any interesting phenomena (see figure12). We analysed the parameters for this case and found that now we came back to the region where $\xi \sim 1$, so that stochastic dynamics came back and harmonics generation is supressed.

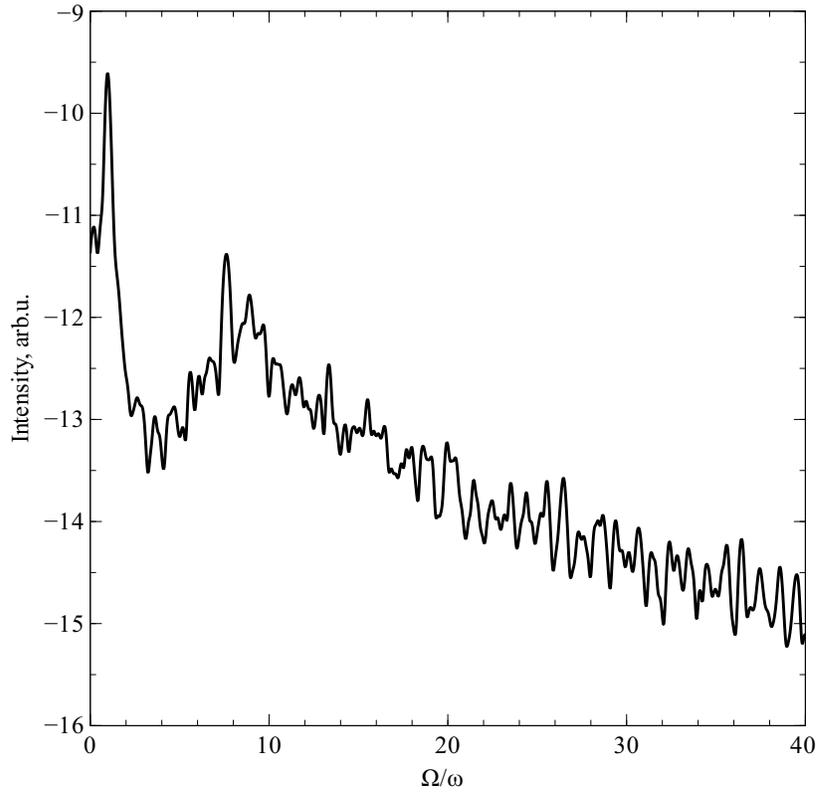

Figure 10: Harmonic spectra for the 1 nm 2 times per ion ionized film, the laser field strength is 2 a.u., the pulse duration is 30 fsec, the total MD integrationl time is 100 fsec. Resonant enhancement of 7-th and 9-th harmonics are clearly seen.

## 4 Conclusion and outlook

Harmonics generation may only be effective if different electrons oscillate in phase. Earlier we considered various situations but did not stressed that the harmonics generation process is a collective classicaly coherent effect. Though it is almost clear, to prove the statement above we now show typical harmonic spectra generated by three different electrons. On the figure 13 we present these spectra for the 1 nm 2 times ionized film, the 2 a.u. laser field with the pulse duration 25 fsec, and the total MD integration time 100 fsec. This situation corresponds to the totak spectrum shown on the figure 10. It is seen, that the spectra of the individual electrons are very wide, and even the laser frequency is not very much notable. However, the spectra generated by the system, presented on figure 10, have much less dispersion and enhancements in certain parts.

After consideration of different aspects of the problem of harmonics generation in an ultrathin nanofilm we now can divide the laser-film interaction scenario at the considered region of parameters into three stages, as it is shown on figure 14. At the beginning of the laser pulse when the electrons are not yet heated one part of the electrons is bound to the film and oscillates within a small region around the ionic core. Another part of the electrons oscillates with a great amplitude, but thermal motion of the electrons in this part is also small. This means that the electron cloud moves to some extend more as a single whole. At this stage (shown in red on the figure 14) perturbation theory gives relevant results, but does not reproduce even harmonics in the spectrum. Later during the interaction the laser pulse heats the electrons, and a part of them evaporates. This destroys the electron cloud, and only collective oscillations of the order of the plasma frequency then survive. The plasma frequency decreases during the interaction process because when electrons are heated they have decreased density. When the plasma frequency decreases to a sertain value, it may be resonantly amplified by a pump laser, more likely at the end of the pulse, if it is few tens of femtoseconds (shown in green on figure 14). Finally, when the laser pulse has gone, there may be residual oscillations in the film with the frequency of the order of the plasma frequency at the end of the pulse (shown in blue on the figure 14). This residual oscillations may generate harmonics for a long time, which is defined by either the energy losses of the excited plasma or the Coulomb explosion of the ionized film. It may happen with some special conditions, that film is 'prepared' before the main pulse comes in such a way, that the plasma oscillations are already excited. In this case we found the effective energy pumping from an incident laser pulse to the plasma oscillations, as it is shown on the time-resolved spectrum on figure 15. During the pulse even harmonics of the plasma frequency are generated, which corresponds to the above mentioned compressibility of the electron gas. When the laser pulse is gone, the plasma oscillations remain with the much higher intensity and also odd harmonics of the plasma frequency are generated because of the nonlinearity of the system. This strong residual plasma oscillations radiate until the film is Coulomb exploded. Note that in the case of a macroscopic square of a film the Coulomb explosion may be suppressed because of the free metal electrons which may partly fill the ionized spot and reduce its positive charge.

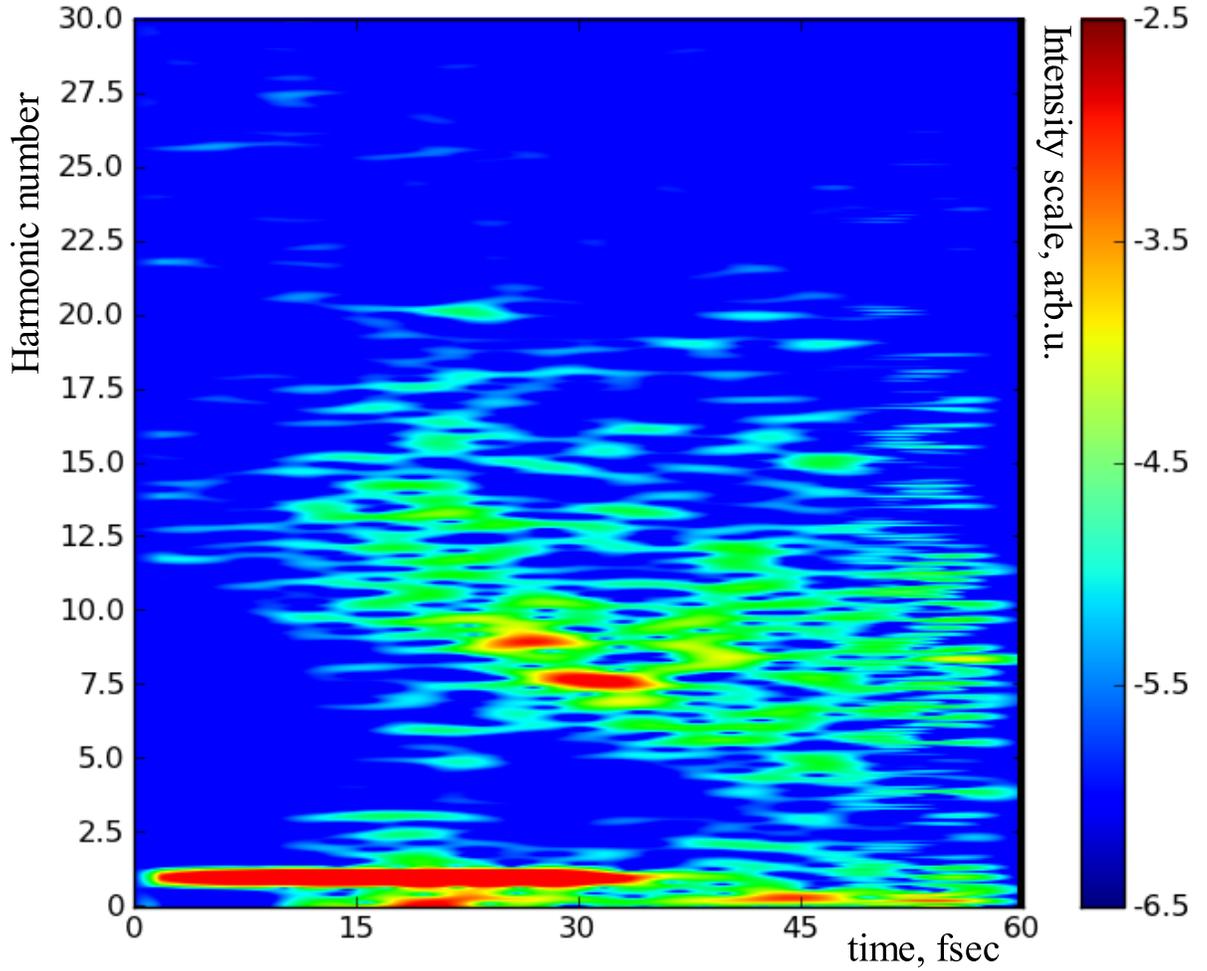

Figure 11: The same as on figure 10, but the spectrum is resolved in time. The decrease of the plasma oscilation frequency and the resonant enhancement of 7-th and 9-th harmonics at the end of the laser pulse are seen.

Finally, we analysed the possibility of harmonics generation in clusters and in ultrathin nanofilms at the regime of a very large electrons oscillation amplitude compared to the nanotarget diameter. We found that in clusters there no such parameters exist to enter the mentioned regime, while in ultrathin nanofilms this is possible. Ultrathin nanofilms were showed to be an effective and perspective source of harmonics generation when irradiated by a high intensity, but not relativistic pump laser field. Strong long-range potential of a charged film effectively bounds electrons and may result in different new phenomena in a laser-nanotarget interactions.

We greatly appreciate productive discussions with S. V. Popruzhenko, and S. P. Goreslavsky. We are sincerely grateful to S. Kelvich for the technical help. This work was supported in part by the Deutsche Forschungsgemeinschaft (SM 292/1-1), RFBR (09-02-00773-a) and the Russian Federal Program RGP (contract No P1546).

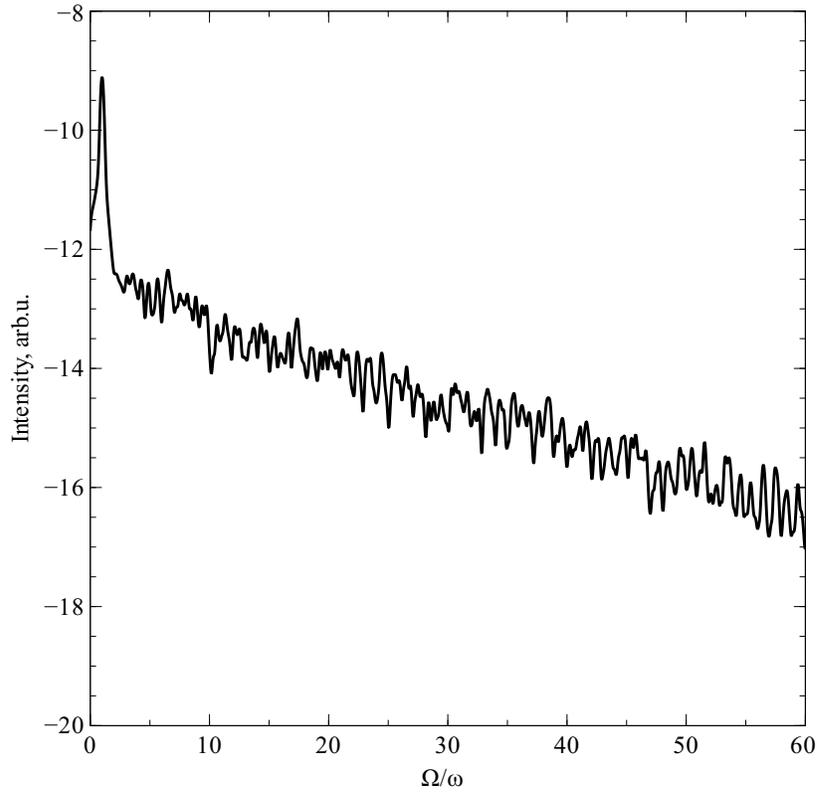

Figure 12: The same as on figure 10, but for the 4 times per ion ionized film.

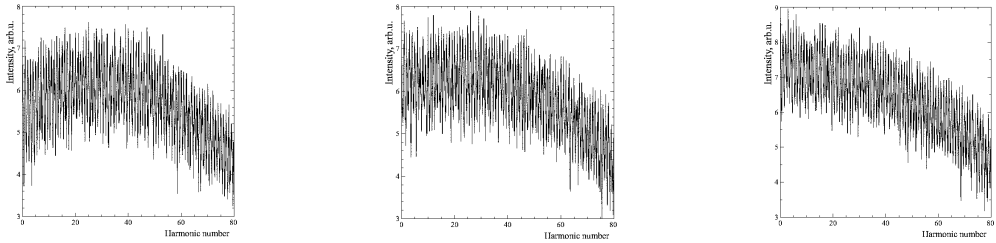

Figure 13: Harmonic spectra for 3 individual electrons, 2 times per ion ionized 1 nm film, the field strength is 2 a.u., the pulse duration is 25fsec, the total MD integration time is 100 fsec.

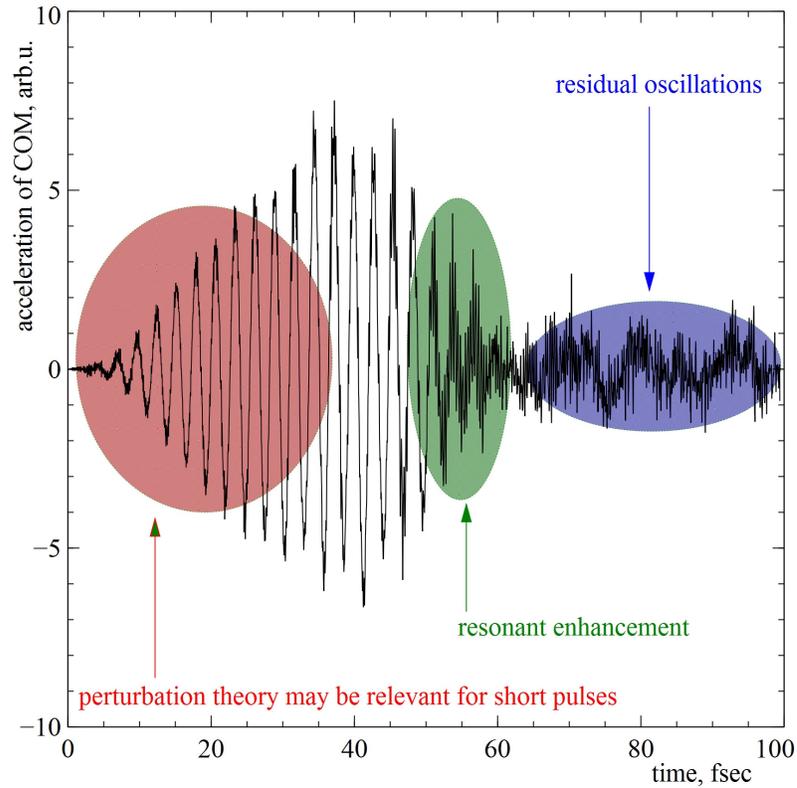

Figure 14: Second derivative of the dipole moment of the whole film for the 1 nm 1 times ionized film, the laser field strength is 2 a.u., the pulse duration is 25 fsec, total MD integration time is 100 fsec

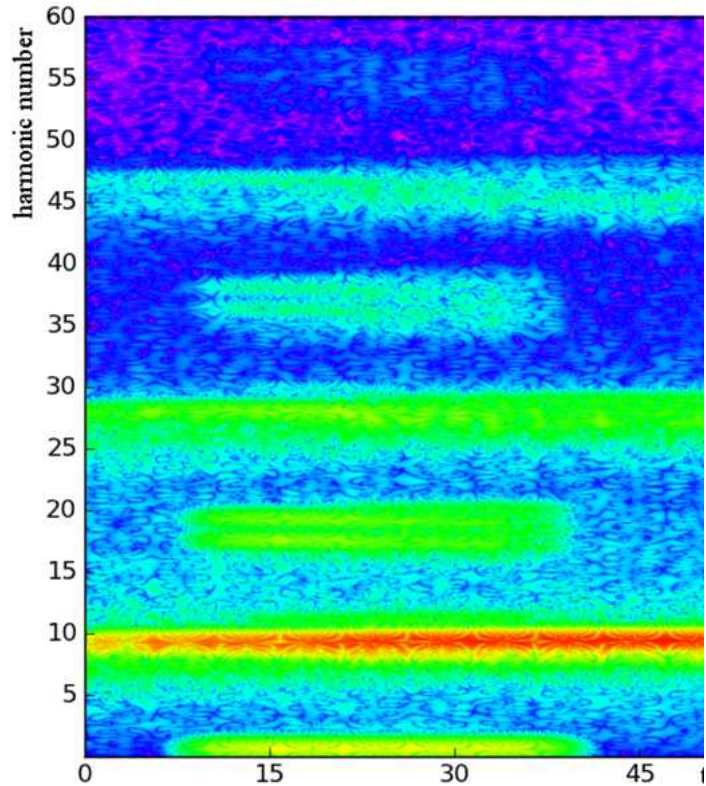

Figure 15: Time-resolved harmonic spectra for a film with initially excited plasma oscillations. The filmwidth is 1nm, the field strength is 2 a.u., the pulse duration is 25 fsec, the total integration time is 60 fsec. The plasma oscillations are amplified by 1.5 orders of magnitude, and become even greater, than the pump laser intensity.